%% file: main.tex
\documentclass[aip,amsmath,amssymb,reprint]{revtex4-1}

\usepackage{graphicx}
\usepackage{dcolumn}
\usepackage{bm}
\usepackage{epstopdf}
\usepackage{xcolor}
\usepackage{natbib}
\usepackage{subcaption} 
\usepackage{float}
\usepackage{tabularx}
\usepackage{amsmath}
\usepackage{mathrsfs}
\usepackage[a]{esvect}
\setcitestyle{square}
\usepackage[toc,page]{appendix}
\usepackage[font={small,normalfont}]{caption}
\usepackage[colorlinks=true, allcolors=blue]{hyperref}
\usepackage[normalem]{ulem}

\graphicspath{{./figures/}}

\newcommand\redsout{\bgroup\markoverwith{\textcolor{black}{\rule[0.5ex]{2pt}{0.4pt}}}\ULon}

\begin{document}

\title{Impact of ion dynamics on laser-driven electron acceleration and gamma-ray emission in structured targets at ultra-high laser intensities}

\author{Tao Wang}
\affiliation{Center for Energy Research, University of California at San Diego,\\ La Jolla, CA, 92093, USA}
\affiliation{Department of Mechanical and Aerospace Engineering, University of California at San Diego,\\La Jolla,
CA 92093, USA}

\author{Zheng Gong}
\affiliation{SKLNPT, School of Physics, Peking University, Beijing 100871, China}
\affiliation{Center for High Energy Density Science, The University of Texas, Austin, TX 78712, USA}

\author{Katherine Chin}
\affiliation{Department of Mechanical and Aerospace Engineering, University of California at San Diego,\\La Jolla, CA 92093, USA}

\author{Alexey Arefiev}
\affiliation{Center for Energy Research, University of California at San Diego,\\ La Jolla, CA, 92093, USA}
\affiliation{Department of Mechanical and Aerospace Engineering, University of California at San Diego,\\La Jolla,
CA 92093, USA}

\date{\today}

\begin{abstract}
We examine the impact of the ion dynamics on laser-driven electron acceleration in an initially empty channel irradiated by an ultra-high intensity laser pulse with $I > 10^{22}$ W/cm$^2$. The negative charge of the accelerated electrons inside the channel generates a quasi-static transverse electric field that causes gradual ion expansion into the channel. Once the ions fill the channel, the pinching force from the quasi-static magnetic field generated by the accelerated electrons becomes uncompensated due to the reduction of the quasi-static transverse electric field. As a result there are two distinct populations of accelerated electrons: those that accelerate  ahead  of  the  expanding  ion  front while moving predominantly forward and those that accelerate in the presence of the ions in the channel while performing strong transverse oscillations. The ions diminish the role of the longitudinal laser electric field, making  the  transverse  electric  field  the  dominant contributor to the electron energy. The ion expansion also has a profound impact on the gamma-ray emission, causing it to become volumetrically distributed while reducing the total emitted energy. We formulate a criterion for the laser pulse duration that must be satisfied in order to minimize the undesired effect from the ions and to allow the electrons to remain highly collimated. \textcolor{black}{We demonstrate the predictive capability of this criterion by applying it to assess the impact of a given pre-pulse on ion expansion. Our results provide a} guideline for future experiments at multi-PW laser facilities with ultra-high intensities.
\end{abstract}

\maketitle

\section{Introduction}

Significant technological progress in laser engineering has made it possible to reliably generate ultra-short laser pulses\cite{1985_CPA,Mourou_etal_2006}. Multi-PW laser pulses are now within reach due to the technology that allows one to compress laser pulses to just tens of femtoseconds\cite{ELI_wb,XCELS,Apllon_laser,Vulcan_Laser,Bella_laser}. Technological improvements have also increased the maximum achievable laser intensity and thus the maximum strength of electric and magnetic fields. Such laser pulses offer a unique opportunity to probe high-intensity light-matter interactions in laboratory conditions. These pulses have also been suggested as drivers for a wide range of applications that require beams of energetic particles\cite{esarey_LWFA_RMP,borghesi_2002_proton_radiography} and radiation\cite{bulanov2011design,di2012extremely,AGRThomas_radiation_review}.

The coupling of the laser energy to the irradiated material is the key for most applications. At high laser intensities, a laser-matter interaction quickly becomes an interaction of a laser pulse with a plasma. The laser fields at the leading edge of a high-intensity laser pulse are already strong enough to ionize the irradiated material. The laser pulse transfers its energy to the plasma electrons and then this energy can be converted into kinetic energy of other particles, such as ions~\cite{macchi_2013_RMP,Arefiev_NJP_2016,Daido_review_ion,bin2015ion,Murakami_ScRep_2018,bin_2018_ion_enhanced,arefiev2018_ion_shotgun}, or into x-rays~\cite{corde_2013_RMP} and gamma-rays~\cite{ridgers2012dense,brady2012laser,vranic2014all,ji2014radiation,RRD_bulanov2015,huang2016_pre,Stark2016_PRL,gonoskov2017_prx,Oliver_pair,huang2018tabletop}. This is the reason why generation of energetic electrons is a critical component even for applications that focus on secondary particle and radiation sources.

There are multiple approaches to effective generation of energetic electrons~\cite{tajima1979,esarey_LWFA_RMP,pukhov1999_DLA,pukhov2002strong,arefiev2012_PRL,arefiev2016beyond_DLA,gong2018_FSSA,vranic2018extremely}. \textcolor{black}{We are specifically interested in the regime of \textit{direct laser acceleration} where the energy is directly transferred from the laser electric fields to irradiated electrons. The distinctive feature of this regime is its ability to generate a large population of energetic electrons with a significant energy spread, needed for generation of secondary particle and radiation beams.} The energy transfer is proportional to $({\bf{E}} \cdot {\bf{v}})$, where \textbf{E} is the laser electric field and \textbf{v} is the electron velocity. Conceptually, there are two possibilities for achieving the energy transfer. One possibility is to employ the electric field transverse to the laser propagation, $E_{\perp}$ which is the dominant component for very wide laser beams. However, most high-intensity laser pulses are tightly focused, which introduces a non-negligible longitudinal electric field, $E_{\parallel}$\cite{Gong_hollow_channel}. This allows a second option for the energy transfer from the laser pulse to become available, with $E_{\parallel}$ doing most of the work~\cite{kluge2012_NJP,xiao2016_PRE,Willingale_2018NJP}.


The option of leveraging the longitudinal laser field is particularly attractive if the end-goal is to generate a highly relativistic electron beam with low angular divergence. An important point is that a forward moving electron can continue effectively gaining energy from $E_{\parallel}$, whereas an effective energy transfer from $E_{\perp}$ requires transverse oscillations that cause finite divergence of the accelerated electrons. This can be illustrated by considering the work $W \propto ({\bf{E}} \cdot {\bf{v}})$ for a relativistic electron moving at an angle $\theta \ll 1$ to the direction of the laser pulse propagation. The longitudinal and transverse velocity components of the electron can be approximated as $v_{\parallel} \approx c \cos \theta \approx c$ and $v_{\perp} \approx c \sin \theta \approx c \theta$. The work by the transverse laser field decreases with $\theta$, $W_{\perp} \propto E_{\perp} v_{\perp} \propto \theta$. On the other hand, the work by the longitudinal laser electric field is independent of $\theta$ for the electron motion that is close to parallel.

Hollow-core structured targets have been shown to create the conditions necessary for generating highly collimated electron beams through direct acceleration by longitudinal laser electric fields\cite{naumova2004_PRL,kluge2012_NJP,jiang2014effects,jiang2016microengineering,xiao2016_PRE}. A hollow-core target guides a tightly focused laser beam without allowing it to diverge~\cite{ji2016towards,snyder2019relativistic}. This way $E_{\parallel}$ can be enhanced and  maintained over distances much greater than the Rayleigh length, which allows for the electrons to continue gaining energy while moving forward with the laser pulse. The accelerated forward-moving electrons create transverse quasi-static electric and magnetic fields. The resulting transverse force has been shown to be very small for forward moving electrons\cite{Gong_hollow_channel}. As a result, the electrons can continue their forward acceleration without gaining transverse velocity that would increase the electron divergence. 

It must be pointed out that the ions in the walls of the hollow-core region do not experience the same force compensation as the accelerated electrons. They feel the inward pull of the transverse electric field generated by the accelerated electrons, whereas the effect of the magnetic field on them is negligible. This means that nothing prevents ions from filling the hollow region during electron acceleration. However, the disruption to the described force compensation resulting from their positive charge would have significant implications in the context of electron acceleration.

\textcolor{black}{Our goal is to determine a condition that specifies when the ion dynamics in an initially empty channel becomes important and must be taken into consideration. Typically, laser-irradiated structured targets are studied using particle-in-cell (PIC) simulations~\cite{naumova2004_PRL,jiang2014effects,ji2016towards,micro_channel_2017sci_reports}. In this approach, the ion dynamics is treated self-consistently, but an extensive parameter scan is required to find the sensitivity on laser pulse and target parameters. The knowledge of the parametric dependence might be particularly important in those cases when the ionization state for the high-Z target material is prescribed~\cite{ji2016towards,preionized_multihole_target,yu2018generation}, rather than being evaluated as a part of the kinetic simulation. }

\textcolor{black}{An additional motivation for this work is the need to assess the impact of a given pre-pulse on ion expansion. It is common to disregard laser pre-pulse in simulations of structured targets, because a full-scale PIC simulation including even just a ps-long pre-pulse is much more computationally expensive than a simulation of the main pulse that is only tens of fs in duration. On the other hand, it is important to know the role of a pre-pulse in order to determine whether or not a given setup can be achieved under specific experimental conditions where a pre-pulse is almost always present.}


In this work, we examine the impact that the ion dynamics have on laser-driven electron acceleration in initially hollow-core channels. We find that the time required for the ions to fill the channel is reduced as the laser pulse amplitude increases. Consequentially, the longitudinal electron acceleration by a relatively short but very intense laser pulse can be disrupted. For example, this is the case for a 35 fs pulse with a peak intensity of $5 \times 10^{22}$ W/cm$^2$ propagating though a $\mu$m wide channel. We also find that when ions are present in the channel, the energy transfer to the electrons is no longer dominated by $E_{\parallel}$ and is primarily due to $E_{\perp}$ instead. The increase in the transverse oscillation caused by the presence of ions also has a profound impact on the gamma-ray emission pattern by the accelerated electrons. The emission switches from being concentrated exclusively at the surface of the channel\cite{yu2018generation}, to being volumetrically distributed. 

The manuscript is organized in the following way. In Section~\ref{Sec-2} we estimate the time needed by the ions to fill a channel of a given radius. \textcolor{black}{We formulate a specific criterion for the laser pulse duration that must be satisfied in order to minimize the undesired effect from the ions and to allow the electrons to remain highly collimated.} In Section~\ref{Sec-3}, we examine the impact of the ion mobility on the spectrum of the laser-accelerated electrons. In Sections~\ref{Sec-4} and \ref{Sec-5}, we focus on the time-resolved ion impact on electron acceleration. \textcolor{black}{In Section~\ref{Sec-prepulse}, we demonstrate the predictive capability of the criterion developed in Sec.~\ref{Sec-2} by applying it to assess the impact of a given pre-pulse on ion expansion.} In Section~\ref{Sec-6}, we show that the ion mobility qualitatively changes the photon emission by the laser-accelerated electrons. Finally, in Section~\ref{Sec-7} we summarize our results and discuss the implications of our findings.



\begin{figure}\centering
    \includegraphics[width=0.49\textwidth]{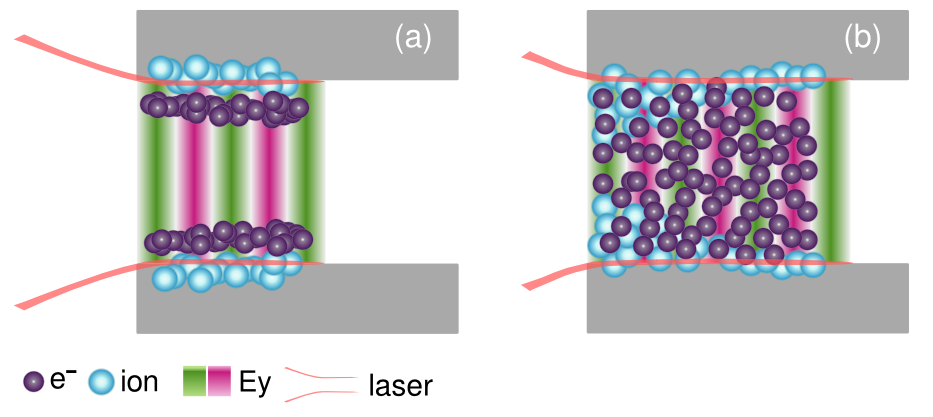}
\caption{\textcolor{black}{Schematic figure for the electron and ion motion inside the channel. The purple and cyan circles represent electrons and ions. The green-red color depicts the laser electric field $E_y$. (a) Electron extraction stage described by Eq.~(\ref{n_*}). (b) Ion expansion stage described by Eq.~(\ref{E_y}).}} \label{schematic_fig}
\end{figure}

\section{Estimates for the ion dynamics in a hollow-core channel} \label{Sec-2}

We are considering a setup where a solid target with a hollow core is irradiated by a high-intensity laser beam. The beam is focused at the channel opening. The focal spot of the beam is assumed to exceed the width of the channel. This condition ensures that the longitudinal component of the laser electric field is enhanced and maintained over a distance that is significantly longer than the Rayleigh length. It has been previously shown that such a field can generate energetic electrons with very low angular divergence~\cite{Gong_hollow_channel}, provided that the ion motion can be neglected.

The accelerated electrons are extracted from the channel walls and injected into the channel by an oscillating transverse electric field of the laser beam. The charge of the injected electrons generates a quasi-static transverse electric field~\cite{Gong_hollow_channel}. This field is directed inwards from the walls toward the central axis of the channel, causing the ions exposed to the field to be pulled into the channel. We are interested in determining when the inward ion motion starts to affect the electron acceleration.

In the context of the transverse ion motion, the key time scale is the time it takes for ions to reach the central axis of the channel through the process shown in Fig. \ref{schematic_fig}. In order to estimate this time, we use a simple model where the electron population is approximated by a uniform negatively charged cloud whose width is equal to the initial width of the channel. The quasi-static transverse electric field generated by the electrons is then given by
\begin{equation} \label{E_y}
    E_y = - 4 \pi |e| n_e y,
\end{equation}
where $n_e$ is the electron density, $e$ is the electron charge, and $y$ is the transverse coordinate. The system of coordinates is aligned such that we have $y = 0$ along the central axis of the channel. The equation of motion for a non-relativistic ion with mass $m_i$ and charge state $Z$ in this field reads
\begin{equation}
    \frac{d v_i}{dt} = \frac{Z |e|}{m_i} E_y,
\end{equation}
where 
\begin{equation} \label{v_y}
    v_i = dy / dt
\end{equation}
is the ion velocity. Equations~(\ref{E_y}) - (\ref{v_y}) fully describe the dynamics of the innermost ions under the given approximations.

From Eqs.~(\ref{E_y}) - (\ref{v_y}) we directly find an expression for the ion velocity as a function of $y$:
\begin{equation} \label{v_i}
    v_i^2 = \frac{Z m_e}{m_i} \omega_{pe}^2 \left(R^2 - y^2 \right),
\end{equation}
where $R$ is the distance from the channel axis to the wall, $\omega_{pe} = \sqrt{4 \pi n_e e^2 / m_e}$ is the electron plasma frequency, and $m_e$ is the electron mass. We have introduced $\omega_{pe}$ in order to make it easier to express our final result in terms of the laser pulse amplitude. Using Eqs.~(\ref{v_i}) and (\ref{v_y}), we now find that the innermost ions reach the axis of the channel after a time interval 
\begin{equation} \label{Delta t}
    \Delta t = \frac{\pi}{2} \sqrt{\frac{m_p}{m_e} \frac{\mu}{\omega_{pe}^2}}
\end{equation}
where $\mu$ is a normalized ion mass-to-charge ratio, 
\begin{equation} \label{mu}
    \mu \equiv \frac{\left. m_i \right/ Z |e|}{\left. m_p \right/ |e|} = \frac{m_i}{Z m_p},
\end{equation}
defined in terms of the proton mass $m_p$.

Our last step is to estimate the electron density $n_e$ in the channel. As previously discussed, the transverse laser electric field can extract electrons from the wall of the channel. The extracted electrons and the ions (that at this stage still remain in the wall) create an electric field whose direction is opposite to the laser electric field. The electron extraction continues unimpeded until the electric field generated due to the charge separation becomes comparable to the laser electric field. The extracted electrons are relativistic, so their displacement from the wall over one laser period is roughly the laser wavelength $\lambda$. The density $n_*$ of the extracted electrons during the first laser period after the extraction can then be roughly estimated using the relation
\begin{equation} \label{n_*}
    E_0 \approx 4 \pi |e| n_* \lambda,
\end{equation}
where $E_0$ is the peak amplitude of the laser electric field. It should be pointed out that the extraction only takes place during a fraction of the laser cycle and is not a continuous process~\cite{ji2016towards,Breizman2003, Breizman_PoP_2005}. This aspect will be accounted for by introducing a numerical multiplier $\xi$ in the final expression (\ref{Delta t - 2}) for the ion travel time. We now introduced a normalized laser amplitude $a_0$ defined as
\begin{equation} \label{a_0}
    a_0 \equiv \frac{|e| E_0}{m_e c \omega},
\end{equation}
where $\omega$ is the laser pulse frequency. Combining Eqs.~(\ref{n_*}) and (\ref{a_0}), we find that the electron plasma frequency corresponding to $n_*$ is approximately given by
\begin{equation} \label{omega_p*}
    \omega_{pe*}^2 \approx a_0 c \omega / \lambda \approx \omega^2 a_0 / 2 \pi.
\end{equation}

Following their injection into the channel, the electrons gradually spread across the channel while moving forward with the laser pulse. We assume that the number of electrons per unit length remains relatively constant during this process. Then, the conservation of particles relates $n_*$ to $n_e$ that we used earlier to estimate the ion dynamics [see Eqs.~(\ref{v_i}) and (\ref{Delta t})]: $n_e R \approx n_* \lambda$. This relation can be rewritten in terms of the electron plasma frequency and, taking into account Eq.~(\ref{omega_p*}), we find that
\begin{equation}
    \omega_{pe}^2 \approx \omega_{pe*}^2 \lambda / R \approx \omega^2 a_0 \lambda / 2 \pi R.
\end{equation}
We substitute this expression for $\omega_{pe}^2$ into Eq.~(\ref{Delta t}) to obtain the time required for the innermost ions to reach the axis of the channel: 
\begin{equation} \label{Delta t - 2}
    \frac{\Delta t}{T} = \xi 
    \sqrt{\frac{\mu}{a_0} \frac{R}{\lambda} \frac{m_p}{m_e}},
\end{equation}
where $T \equiv 2 \pi / \omega$ is the laser period and $\xi$ is a numerical factor. We have made a number of approximations to find how $\Delta t$ scales with $\mu$, $a_0$, and $R$. However, the presented derivation lacks the precision needed to determine $\xi$. In what follows, we will use fully self-consistent kinetic simulations to confirm the obtained scaling and to determine the value of $\xi$ numerically (as shown in Sec.~\ref{Sec-4}, $\xi\approx1.4$).


In our estimates, we have assumed that the ions are not relativistic, i.e. $\gamma_i-1 \ll 1$, where $\gamma_i=1/\sqrt{1-v_i^2/c^2}$. The non-relativistic assumption is equivalent to $v_i^2 \ll c^2$. The ions reach their maximum velocity at the axis of the channel where $y = 0$. It follows from Eqs.~(\ref{v_i}) and (\ref{Delta t}) that the condition $v_i^2 \ll c^2$ is satisfied during the ion motion towards the axis if
\begin{equation} \label{v_i-final}
    \left[\max\left(v_i\right)\right]^2 = \left(\frac{\pi}{2} \frac{R}{\Delta t}\right)^2 \ll c^2.
\end{equation}
This condition is equivalent to
\begin{equation} \label{rel_condition}
    \left(\frac{\Delta t}{T}\right)^2 \gg \left(\frac{\pi}{2} \frac{R}{\lambda}\right)^2.
\end{equation}
If the value of $\Delta t$ given by Eq.~(\ref{Delta t - 2}) fails to satisfy the condition (\ref{rel_condition}), then relativistic corrections must be included. These corrections increase the ion travel time to ensure that $\Delta t > R/c$ in agreement with the relativistic mechanics.

We conclude this section by using the derived expressions to estimate the ion travel time $\Delta t$ for a laser pulse with a peak normalized amplitude of $a_0 \approx 200$. This is the amplitude projected for tightly focused laser beams of the ELI laser facilities~\cite{ELI_wb}. In the case of a channel with $R \approx 3\lambda$, we have
\begin{equation}
    \Delta t / T \approx \xi \sqrt{30 \mu}.
\end{equation}
If the target material is fully ionized carbon, then $\mu \approx 2$ and 
$\Delta t / T \approx 8 \xi$. Assuming that $\xi$ is of the order of unity, we find that it takes only several laser periods for the ions to fill the channel. The condition (\ref{rel_condition}) is still satisfied for these extreme parameters, so the ions are well described using the non-relativistic equations of motion.

The estimates presented in this section show that the number of electrons injected into the hollow channel and their density increase with laser amplitude $a_0$. These electrons create a transverse electric field that causes ions to be injected into the channel as well. Our main conclusion is that the ion injection can take place on a time scale \textcolor{black}{shorter than} the duration of a laser pulse of ultra-high intensity.


\begin{figure*}[htb]
\centering
	\begin{minipage}[c]{0.95\textwidth}{
		\includegraphics[width=\linewidth,trim={0.2cm 1.3cm 0.5cm 1cm},clip]{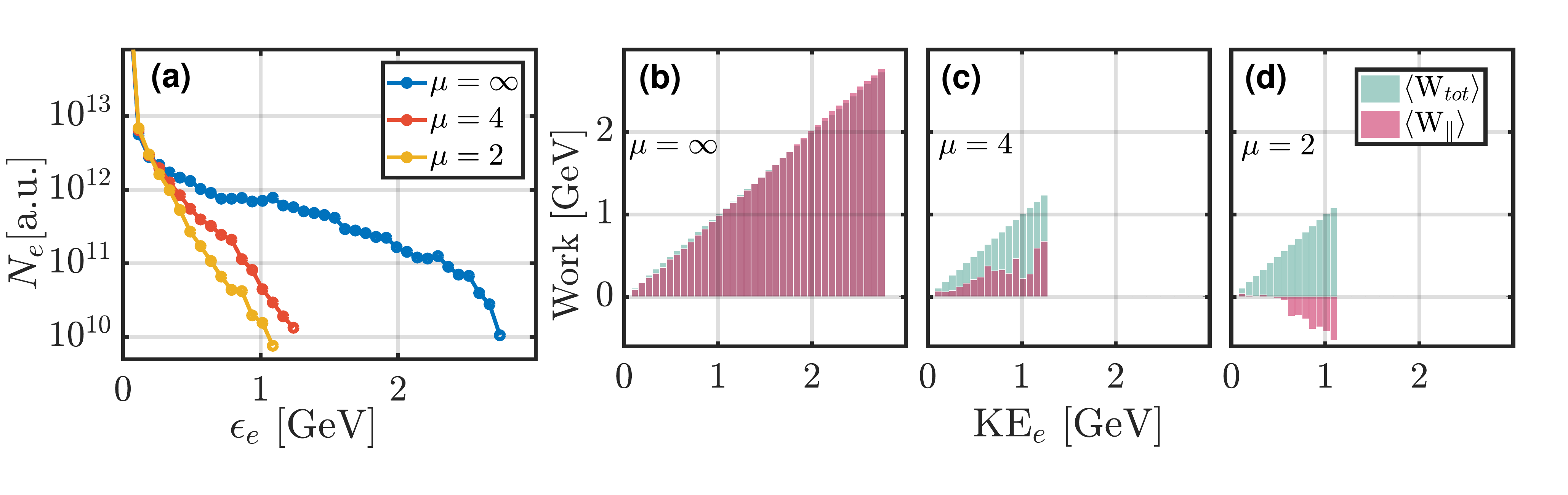}}
	\end{minipage}%
		\caption{Electron energy spectra (a) and the corresponding contributions to the electron kinetic energy $\mathrm{KE}_e$ by the longitudinal electric field (b,c, and d) for three different values of the normalized ion mass-to-charge ratios $\mu = 2$, 4, and $\infty$. All of the snapshots are taken at $t = 124$ fs, where $t = 0$ fs is the time when the pulse reaches its peak intensity at $x = 0$ $\mu$m in the absence of the target. The bar width in (b) - (d) is 75 MeV. The electron spectrum for $\mu = 2$ saturates at $t > 124$ fs.}
	\label{work_energy}
\end{figure*}

\section{Impact of ion mobility on electron energy spectrum} \label{Sec-3}

The estimates of Sec.~\ref{Sec-2} indicate that the ion motion should become important in narrow hollow-core channels at ultra-high laser intensities with $a_0 > 100$, such as those projected for the ELI laser facilities~\cite{ELI_wb}, even if the laser pulse is only tens of cycles long. In this section, we examine the impact of the ion mobility on the laser accelerated electrons.

The ionization of the target material is the process that directly impacts the ion dynamics by increasing the ion charge state. There are multiple ionization mechanisms that can play a role in the considered setup, but the field ionization is the one primarily responsible for significantly increasing the ion charge state at ultra-high laser intensities. As the ion charge state $Z$ is increased, the mass-to-charge ratio $\mu$ [see Eq.~(\ref{mu})] is reduced. The latter is the parameter that determines the ion response to applied electric fields, with the ions being more mobile for lower mass-to-charge ratios.

In order to investigate the impact of the ion mass-to-charge ratio $\mu$ on electron acceleration, we treat $\mu$ as an input parameter rather than calculating it self-consistently. This approach allows us to keep all of the laser pulse parameters fixed while varying the value of $\mu$. In a fully self-consistent analysis, the laser amplitude $a_0$ impacts the electron acceleration in two ways: directly through electron interaction with the laser electric fields and indirectly by influencing the ion dynamics through the ion mass-to-charge ratio, as described earlier. It is therefore necessary to decouple $\mu$ from $a_0$ and other laser pulse parameters in order to identify the impact of the ions.

We simulate the electron acceleration using a fully relativistic two-dimensional (2D) particle-in-cell (PIC) simulation. The target is a uniform plasma rectangle with a straight empty channel. It is irradiated by a laser pulse focused at the channel entrance. The laser parameters are similar to those projected for the ELI facility~\cite{ELI_wb}. The system of coordinates is set such that the laser pulse propagates in the positive direction along the $x$-axis. The central axis of the laser pulse corresponds to $y = 0$, while the channel opening is located at $x = 0$. Table~\ref{table_PIC} provides additional simulation details. Note that the electron density is given in terms of the critical electron density $n_{crit}$ defined by the condition $\omega_{pe}^2 = \omega^2$, where $\omega_{pe}^2 = 4 \pi n_{crit} e^2 / m_e$.

\begin{table}[htbp]
\caption{\textbf{Parameters of the 2D PIC simulations}}
\label{table_PIC}
\begin{tabular}{ |l|l| }
  \hline
  \multicolumn{2}{|c|}{2D PIC simulation parameters} \\
  \hline
  \multicolumn{2}{|c|}{}\\
  \multicolumn{2}{|l|}{\textbf{Laser pulse:} }\\
  Peak intensity & $5 \times 10^{22}$ W/cm$^2$ \\
  Normalized field amplitude & $a_0 = 191$ \\
  Wavelength & $\lambda = 1.0$ $\mu$m \\
  Pulse duration & \\
  (FHWM for intensity) & 35 fs\\
  Focal spot size &  \\
  (FWHM for intensity) & 3.9 $\mu$m \\
  Location of the focal plane & $x = 0$ $\mu$m\\
  \multicolumn{2}{|c|}{}\\
  \multicolumn{2}{|l|}{\textbf{Structured target:} }\\
  Total channel width  & $2R = 5.4$ $\mu$m \\
  Location of the channel opening & $x = 0$ $\mu$m\\
  Electron density & $n_e = 100n_{crit}$ \\
  Ion mass-to-charge ratio & $\mu = 2;$ 4; and $\infty$ \\
  Ion charge state & $Z = 6$ \\
  Ion density & $n_i = n_e / Z$ \\
  \multicolumn{2}{|c|}{}\\
  \multicolumn{2}{|l|}{\textbf{General parameters:} }\\
  Target length & 72.0 $\mu$m\\
  Target width & 16.0 $\mu$m\\
  Spatial resolution & $100/\mu m\times 100/\mu m$ \\
  Number of macro-particles per cell  &  \\
  for electrons & 80 \\
  for ions & 40 \\
  \hline
\end{tabular}
\end{table}

Figure~\ref{work_energy}a shows the electron energy spectra for three different ion \textcolor{black}{mass-to-charge} ratios $\mu = 2;$ 4; and $\infty$. We ran three independent 2D PIC simulations to obtain these spectra. Ion mobility is shown to have a negative impact on electron acceleration, since the number of energetic electrons visibly decreases with the reduction of the mass-to-charge ratio. The electron energy $\epsilon_e$ is defined as $\epsilon_e \equiv \gamma m_e c^2$, where $\gamma$ is the electron relativistic factor. All three snapshots of the electron spectra are taken at $t = 124$ fs. The electron spectrum for $\mu = 2$ saturates at $t > 124$ fs. We define $t = 0$ as the time instant when the laser pulse reaches its peak intensity at $x = 0$ in the absence of the target. 

The accelerated electrons can gain their energy from longitudinal $(E_\parallel = E_x)$ and transverse $(E_{\perp} = E_y)$ electric fields, so we separately examine the work performed by each of the components to better understand the observed changes in the electron spectra. The total work performed by the electric fields acting on a given electron during the simulation,
\begin{equation}
\mathrm{W}_{tot} (t) = - |e| \int_{-\infty}^t \left( {\bf{E}} \cdot {\bf{v}} \right) dt',
\end{equation}
is a sum of the work performed by $E_\parallel$ and $E_{\perp}$:
\begin{equation} \label{W_x and W_y}
\mathrm{W}_{tot} = \mathrm{W}_{\parallel} + \mathrm{W}_{\perp} \equiv -|e| \int_{-\infty}^t ({E_\parallel} {v_\parallel} + {E_\perp} {v_\perp}) dt'. 
\end{equation}
In our simulations the electrons are initially cold with $\epsilon_e = m_e c^2$, so that their kinetic energy $\mathrm{KE}_e$ is simply equal to the total work $\mathrm{W}_{tot}$:
\begin{equation}
    \mathrm{KE}_e \equiv \epsilon_e - m_e c^2 = \mathrm{W}_{tot} = \mathrm{W}_{\parallel} + \mathrm{W}_{\perp}.
\end{equation}

Figures~\ref{work_energy}b, \ref{work_energy}c, and \ref{work_energy}d show the contribution from $E_{\parallel}$ to the total work as a function of the electron kinetic energy. Each vertical bar represents a group of electrons whose kinetic energy $\mathrm{KE}_e$ is in the range determined by the width of the bar, $\delta \mathrm{KE}_e = 75$ MeV. $\left< \mathrm{W}_{tot} \right>$ and $\left< \mathrm{W}_{\parallel} \right>$ are $\mathrm{W}_{tot}$ and $\mathrm{W}_{\parallel}$ averaged over the electrons in each of the groups. The contribution from the longitudinal electric field decreases as we decrease $\mu$ from $\infty$ to 4 and it becomes negative for $\mu = 2$.

In the case of immobile ions ($\mu = \infty$), the accelerated electrons gain most of the energy from a longitudinal electric field. Specifically, we find that $\sum \mathrm{W}_{\parallel} / \sum \mathrm{KE}_e \approx 0.97$, where the summation is performed over the electrons with $\epsilon_e \geq 150$ MeV. This agrees with our previously published work~\cite{Gong_hollow_channel} where this field was identified as the longitudinal field of the laser pulse. As we increase the ion mobility by reducing $\mu$ from $\infty$ to 4, the contribution from the longitudinal electric field becomes significantly reduced, with $\sum \mathrm{W}_{\parallel} / \sum \mathrm{KE}_e \approx 0.39$ for $\epsilon_e \geq 150$ MeV. At an even lower value of the mass-to-charge ratio the longitudinal electric field becomes detrimental, as its contribution becomes negative, with $\sum \mathrm{W}_{\parallel} / \sum \mathrm{KE}_e \approx -0.01$ for $\epsilon_e \geq 150$ MeV at $\mu = 2$. 

The results of this section indicate that the increase in ion mobility not only decreases the number of energetic electrons, but it also reduces the contribution to the electron acceleration from the longitudinal laser electric field.


\section{Time-resolved ion impact on electron acceleration} \label{Sec-4}

In Section~\ref{Sec-3}, we examined how ``macroscopic'' characteristics of the accelerated electrons change with the ion mass-to-charge ratio. Since ion expansion into the channel takes time to manifest itself, one might expect  electrons travelling at the back of the laser pulse to be influenced by the ions differently than electrons travelling at the front of the pulse. In what follows, we examine how the ion impact on electron acceleration varies in time.

\begin{figure}[htb]
    \begin{center}
       \includegraphics[width=0.48\textwidth,trim={0cm 0.7cm 0cm 1.0cm},clip]{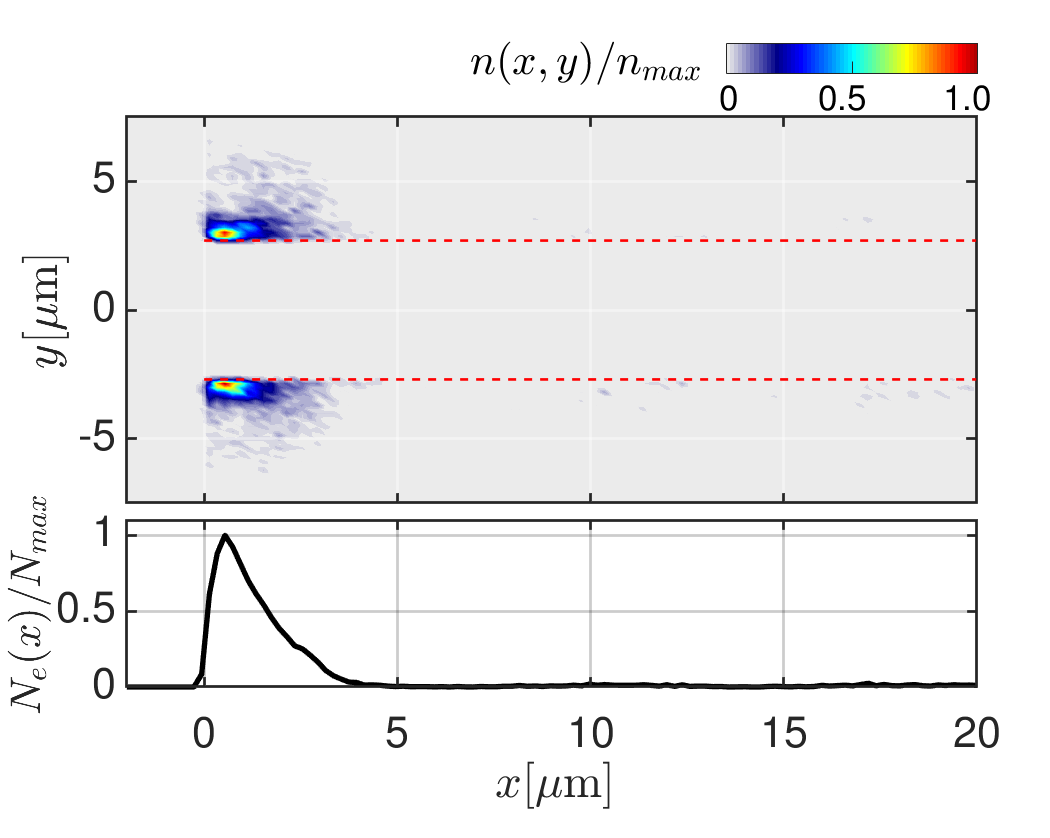}
    \end{center}
\caption{Initial spatial distribution of energetic electrons with $\epsilon_e > 150$ MeV from Fig.~\ref{work_energy} for the case of $\mu = 2$. The curve in panel (b) is the distribution from panel (a) integrated over the transverse coordinate. In both panels the electron numbers are normalized to their maximum values $n_{max}$ and $N_{max}$. The red dashed lines show the initial channel boundary.}
\label{Fig:ini_pos}
\end{figure}

Our first step is to confirm and refine the qualitative picture of electron injection and acceleration presented in Sec.~\ref{Sec-2}. Figure~\ref{Fig:ini_pos} shows the initial positions of the energetic electrons with $\epsilon_e > 150$ MeV at $t = 124$ fs from Fig.~\ref{work_energy} for the case of $\mu = 2$. The majority of the energetic electrons clearly originate close to the channel opening. Additional electron tracking shows that their injection into the channel also takes place in the vicinity of the channel opening (a similar phenomenon has been reported for laser-generated channels in uniform plasma targets~\cite{arefiev2015_JPP,arefiev2016beyond_DLA,robinson2013generating}). The results for $\mu = 4$ and $\mu = \infty$ are omitted, as they are very similar to those shown in Fig.~\ref{Fig:ini_pos}. The main conclusion is that the electron injection into the channel is highly localized. 

\begin{figure*}[htb]
	\begin{subfigure}[c]{.4\linewidth}{
    \includegraphics[width=0.9\textwidth,trim={0cm 0.cm 0cm 0.cm},clip]{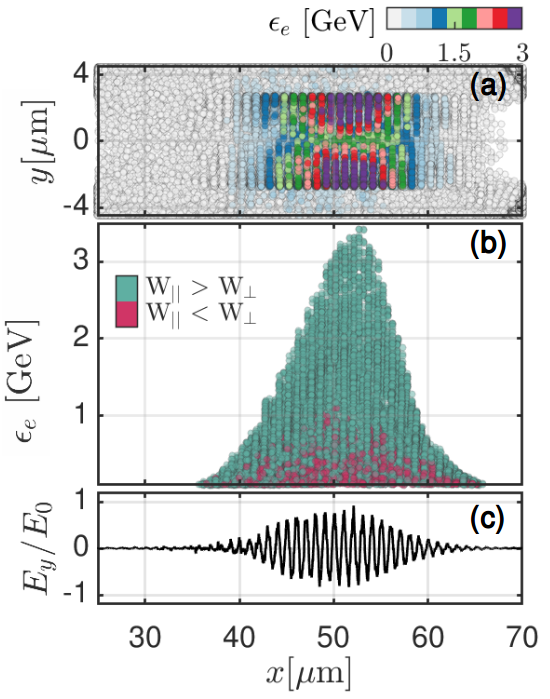}}
    \end{subfigure}%
    \begin{subfigure}[c]{.4\linewidth}{\includegraphics[width=0.9\textwidth,trim={0cm 0cm 0 0cm},clip]{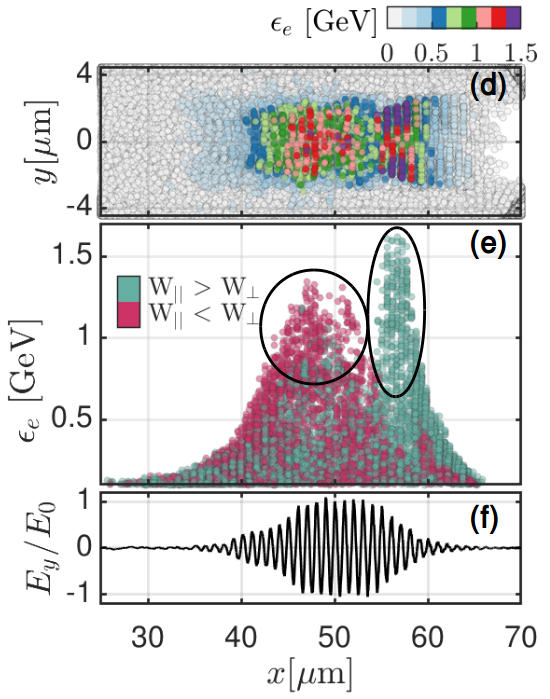}}
    \end{subfigure}
    \caption{Comparison of electron acceleration between channels with mobile ($\mu = 2$) and immobile ($\mu = \infty$) ions in the acceleration region ($x > 7.5$ $\mu$m). The left column corresponds to the immobile case, while the right column corresponds to the mobile case. All of the snapshots are taken at $t = 171$ fs. \textcolor{black}{Panels (a) and (d) show electron position and energy (color-coded). Panels (b) and (e) show the locations and energies of two groups of electrons: electrons with $\mathrm{W}_{\parallel} > \mathrm{W}_{\perp}$ (green) and electrons with $\mathrm{W}_{\parallel} < \mathrm{W}_{\perp}$ (red). In panels (c) and (f),} the line-outs of the transverse electric field are plotted for $y = 0$~$\mu$m and they are normalized to the peak amplitude of the laser electric field $E_0$ in the absence of the target.}
    \label{Fig:two_groups}
\end{figure*}

The simulations also show that the electron energy gain takes place over a distance that significantly exceeds the length of the region where the injection takes place. This agrees well with the qualitative understanding of the electron acceleration. The electrons move forward with the laser pulse and can gain energy as long as they remain in the favorable phase of the laser wave. The spatial period of the laser field is $\lambda$. An electron that moves forward with a highly relativistic longitudinal velocity would travel a distance roughly equal to $\gamma \lambda$ before slipping by one laser wavelength with respect to the moving field pattern of the laser pulse, where $\gamma$ is the characteristic relativistic factor of the electron. We can therefore estimate the acceleration distance as $\Delta x \approx \gamma \lambda$, which yields $\Delta x \approx 10^3 \lambda$ for 500 MeV electrons. This distance exceeds the length of the injection region by at least two orders of magnitude.

The ion dynamics can impact both the electron injection and their subsequent acceleration by the laser pulse. However, the described separation of longitudinal scales allows us to separate these two aspects. In this work, our primary focus is on the ion impact on electron acceleration. We isolate the corresponding physics by utilizing a simulation setup where the ions in the injection region are immobile ($\mu = \infty$). The ion mass-to-charge ratio is only varied further along the length of the channel where the electron acceleration takes places. Specifically, the region with immobile ions is located at $x \leq 7.5$ $\mu$m, whereas the region with variable $\mu$ is located at $x > 7.5$ $\mu$m. In what follows, the former will be referred to as the \textit{injection region}, whereas the later will be referred to as the \textit{acceleration region}.

In order to clearly identify the impact of the ion mobility in the acceleration region, we begin by examining the regime where the ions are immobile. Figures~\ref{Fig:two_groups}a and \ref{Fig:two_groups}b provide detailed information about accelerated electrons for the case with immobile ions ($\mu = \infty$) in the acceleration region. The energetic electrons in Fig.~\ref{Fig:two_groups}a are grouped in well-defined bunches whose periodicity mirrors that of the laser electric field shown in Fig.~\ref{Fig:two_groups}c. It is evident from Fig.~\ref{Fig:two_groups}b that most of the electron energy is gained from the longitudinal electric field. The most energetic electrons are located in the middle of the laser pulse where the laser field has the highest amplitude. This is consistent with our understanding of the electron acceleration mechanism dominated by $E_{\parallel}$, where the energy transfer to a forward moving electrons is directly proportional to the amplitude of the laser electric field. The energetic electrons remain effectively locked in the favorable accelerating stage.

Figures~\ref{Fig:two_groups}d and \ref{Fig:two_groups}e show accelerated electrons in a simulation with mobile ions $(\mu = 2)$ in the acceleration region. The electrons at the front of the laser pulse look very similar to the electrons in the immobile case (see Figs.~\ref{Fig:two_groups}a and \ref{Fig:two_groups}b). However, there is a significant difference in the electrons that follow. Most of their energy is gained from the transverse electric field $E_{\perp}$. The energy of the electrons in this group is reduced compared to the electrons in the immobile case. This result agrees with the result shown in Fig.~\ref{work_energy}, while at the same time it confirms that there is a direct correlation between the energy reduction and the switch from the longitudinally dominated acceleration regime ($\mathrm{W}_{\parallel} > \mathrm{W}_{\perp}$) to that where the role of the transverse electric field is dominant ($\mathrm{W}_{\parallel} < \mathrm{W}_{\perp}$).

\begin{figure*}[htb]
    \centering
    \includegraphics[width=0.76\linewidth,trim={0cm 0cm 0 0cm},clip]{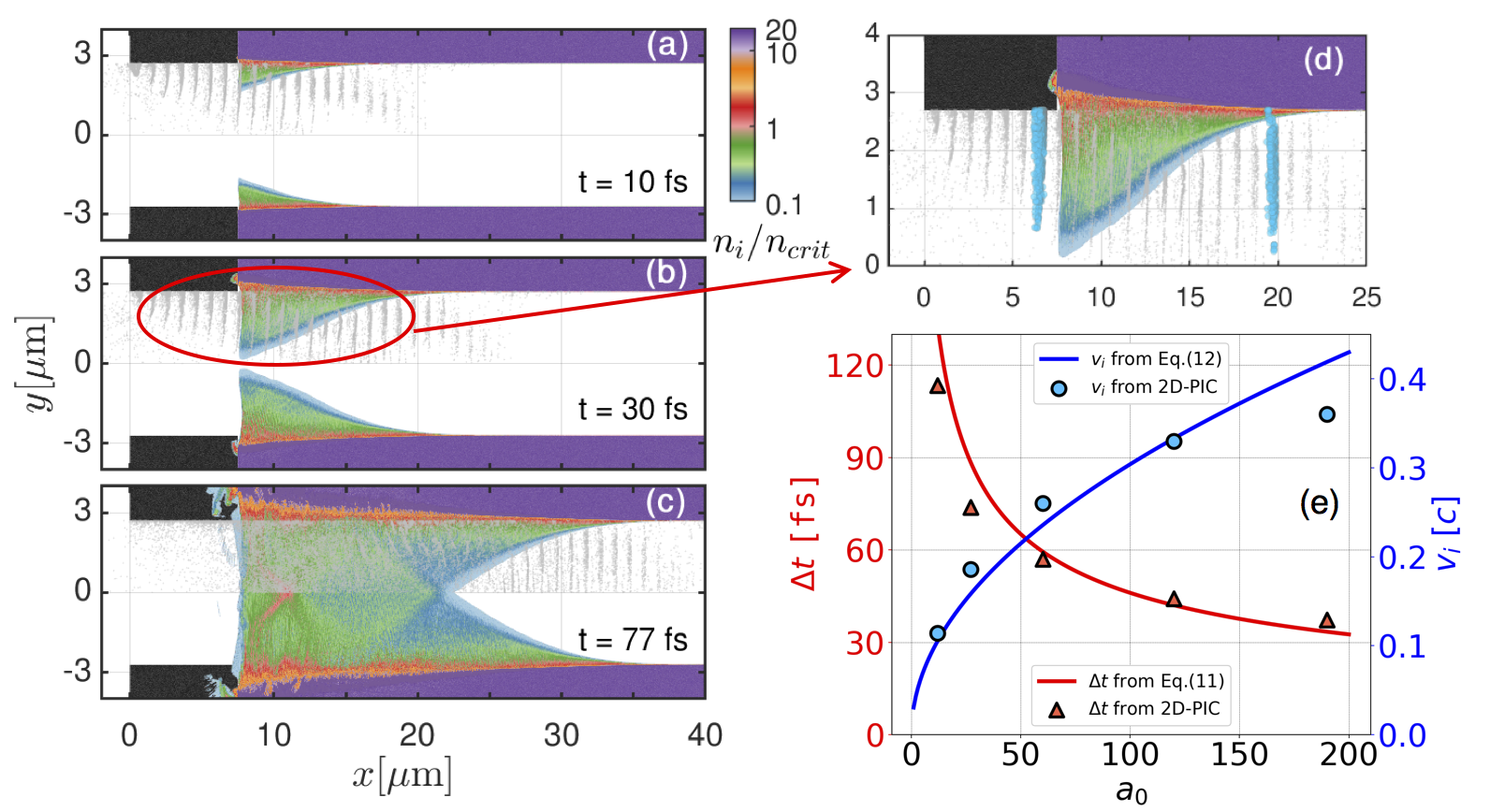}
    \caption{Ion dynamics in a laser irradiated channel with an immobile injection region (shown with dark gray). Panels (a) to (d) show snapshots of the ion density and injected electron bunches (for $y > 0$ $\mu$m only). The two bunches marked with blue in panel (d) are used for detailed tracking in Sec.~\ref{Sec-5}. Panel (e) gives the ion travel time to the central axis and the maximum velocity of the innermost ions as a function of the normalized laser amplitude $a_0$. The curves are the estimates from Sec.~\ref{Sec-2} given by Eqs.~(\ref{Delta t - 2}) and (\ref{v_i-final}) with $\xi \approx 1.4$.} 
    \label{Fig:ion_dynamics_more}
\end{figure*}

We find that the observed change in the electron acceleration is a direct consequence of the ion expansion into the channel. Figures~\ref{Fig:ion_dynamics_more}a, \ref{Fig:ion_dynamics_more}b, and \ref{Fig:ion_dynamics_more}c show snapshots of the ion density together with the snapshots of electron bunches. Note that only electrons located in the upper part of the channel ($y > 0$ $\mu$m) are shown to make the ion density profile more visible. There are bunches that move forward ahead of the expanding ion boundary. The electrons in these bunches remain unaffected by the ions, with $\mathrm{W}_{\parallel} > \mathrm{W}_{\perp}$. Once the ions reach the axis at a given location, the electron bunches that pass through this location afterwards can no longer escape the ion influence (see Fig.~\ref{Fig:ion_dynamics_more}d), i.e. the influence of the ion electric field. These are the electrons with $\mathrm{W}_{\parallel} < \mathrm{W}_{\perp}$. 

We now use our simulation results to determine the time interval $\Delta t$ required for the ions at a given axial location to reach the axis of the channel. For a given axial position $x = x_*$, we define $\Delta t$ as $\Delta t = t_2 - t_1$, where $t_2$ is the time instant when the innermost ions reach the central axis at $x = x_*$ and $t_1$ is the time instant when the laser intensity at $x = x_*$ reaches 10\% of the peak intensity of the laser pulse in the absence of the target. The condition used to define $t_1$ is equivalent to a condition that $|E_y|$ exceeds $0.32 E_0$. We find that the ion travel time $\Delta t \approx 38$ fs for $x_* = 10$ $\mu$m and it is relatively insensitive to the value of $x_*$. By comparing this result with our estimate for $\Delta t$ given by Eq.~(\ref{Delta t - 2}), we find that the free parameter $\xi$ for our laser pulse is $\xi \approx 1.4$.

Figure~\ref{Fig:ion_dynamics_more}e shows the ion travel time as a function of the normalized laser amplitude $a_0$. The triangles are the results of 2D PIC simulations for five different laser amplitudes. There is a good agreement between the scaling given by Eq.~(\ref{rel_condition}) and our numerical results. An important conclusion is that only first 50 fs of an ultra-high intensity laser pulse with $a_0 > 100$ can accelerate electrons in a regime where the role of the longitudinal electric field is dominant and $\mathrm{W}_{\parallel} > \mathrm{W}_{\perp}$.

\begin{figure*}[htb]
    \centering
    \includegraphics[width=0.9\textwidth]{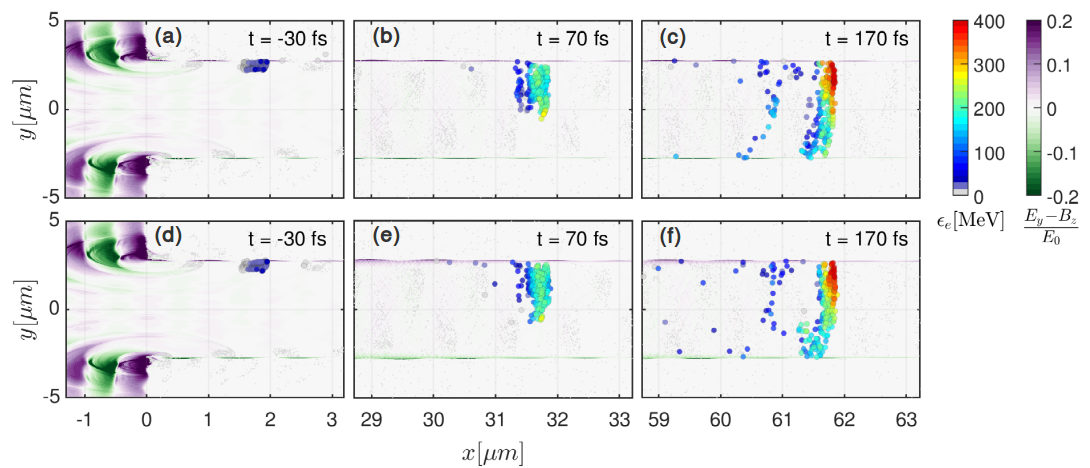}
    \caption{Snapshots of \textbf{instantaneous} fields and an \textit{early bunch} in a channel with immobile, $\mu = \infty$, (top row) and mobile, $\mu = 2$, (bottom row) ions in the acceleration region. The electron energies in the bunch are color-coded. The small gray markers show all energetic electrons with $\epsilon_e > 2$~MeV.}
    \label{Fig:early_bunch_inst}
\end{figure*}

\begin{figure*}[htb]
    \centering
    \includegraphics[width=0.9\textwidth]{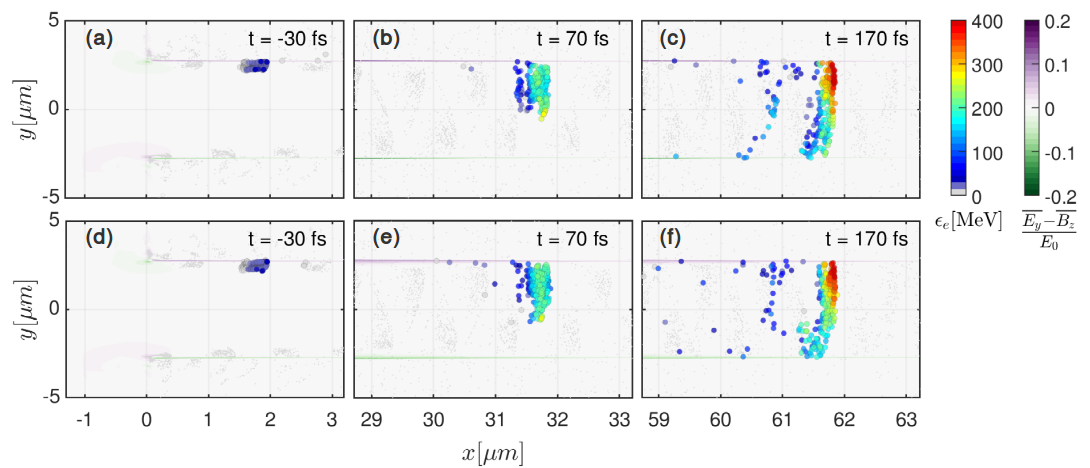}
    \caption{Snapshots of \textbf{time-averaged} fields and an \textit{early bunch} in a channel with immobile, $\mu = \infty$, (top row) and mobile, $\mu = 2$, (bottom row) ions in the acceleration region. The electron energies in the bunch are color-coded. The small gray markers show all energetic electrons with $\epsilon_e > 2$~MeV.}
    \label{Fig:early_bunch_av}
\end{figure*}

Even at the highest considered laser amplitude of $a_0 = 190$ the innermost ions can still be treated as non-relativistic in the considered setup. Their maximum velocity at $a_0 = 190$ reaches $v_i = 0.35 c$, which corresponds to $\gamma_i = (1 - v_i^2/c^2)^{-1/2}\approx 1.07$. The circles in Fig.~\ref{Fig:ion_dynamics_more}e show the maximum velocity of the innermost ions for five different laser amplitudes. These values are obtained from the already mentioned 2D PIC simulations.

The results of this section indicate that there are two distinct populations of accelerated electrons: those that accelerate ahead of the expanding ion front and those that accelerate in the presence of the ions in the channel. The ions reduce the role of the longitudinal laser electric field, making the transverse electric field the dominant contributor to the electron energy ($\mathrm{W}_{\parallel} \ll \mathrm{W}_{\perp}$). 


\section{Individual electron bunch dynamics and the transverse force balance} \label{Sec-5}

\begin{figure*}[htb]
    \centering
    \includegraphics[width=0.9\textwidth]{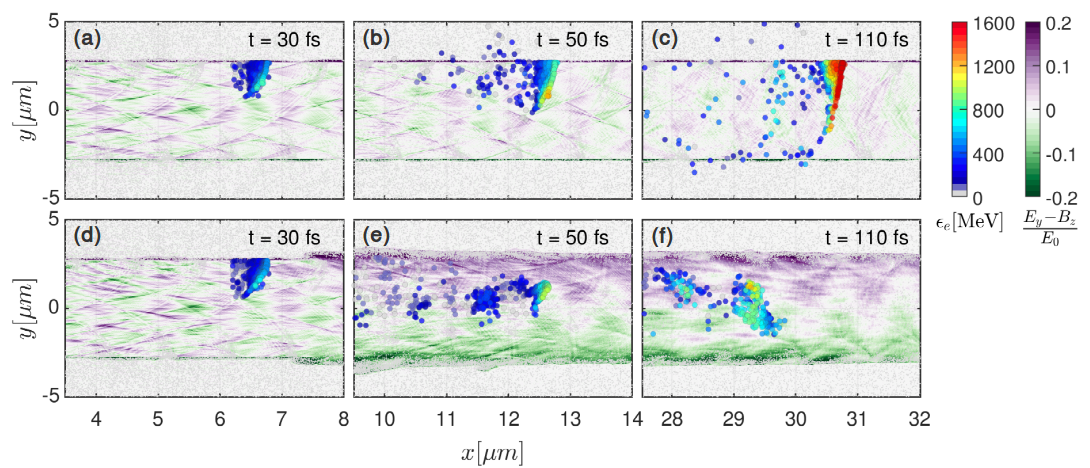}
    \caption{Snapshots of \textbf{instantaneous} fields and a \textit{late bunch} in a channel with immobile, $\mu = \infty$, (top row) and mobile, $\mu = 2$, (bottom row) ions in the acceleration region. The electron energies in the bunch are color-coded. The small gray markers show all energetic electrons with $\epsilon_e > 2$~MeV.}
    \label{Fig:late_bunch_inst}
\end{figure*}

\begin{figure*}[htb]
    \centering
    \includegraphics[width=0.9\textwidth]{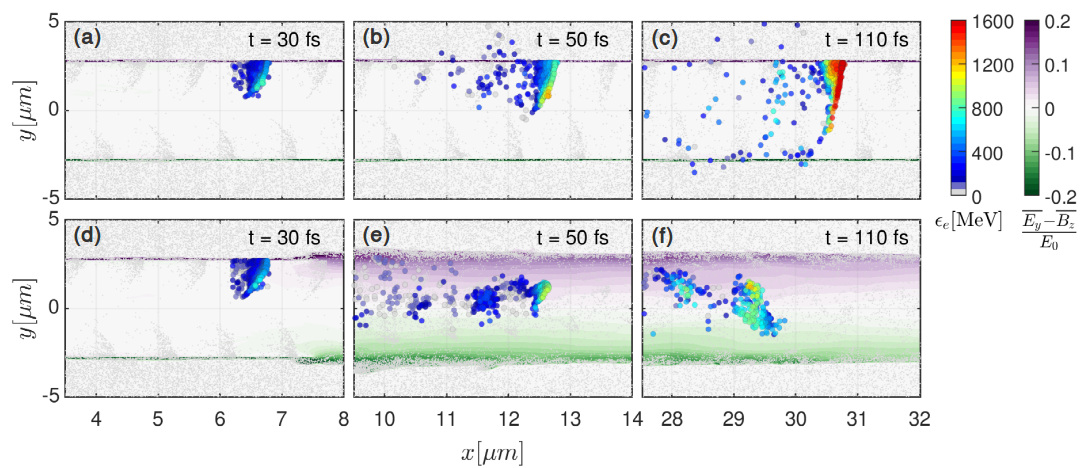}
    \caption{Snapshots of \textbf{time-averaged} fields and a \textit{late bunch} in a channel with immobile, $\mu = \infty$, (top row) and mobile, $\mu = 2$, (bottom row) ions in the acceleration region. The electron energies in the bunch are color-coded. The small gray markers show all energetic electrons with $\epsilon_e > 2$~MeV.}
    \label{Fig:late_bunch_av}
\end{figure*}

The results of Sec.~\ref{Sec-4} indicate that the energy transfer from the transverse electric field to energetic electrons dominates after the ions fill up the channel. However, the amplitude of the transverse laser electric field increases by less than 20\% in the case of mobile ions (see Figs.~\ref{Fig:two_groups}c and \ref{Fig:two_groups}f) and there is no significant decrease in the longitudinal laser electric field. These observations suggest that the dramatic change in the work performed by the laser fields on the accelerated electrons is caused by changes in the electron velocity rather than by changes in the field structure.
In what follows, we examine the time evolution of two electron bunches to understand how the ions dynamics changes the mechanism of electron acceleration.

The two electron bunches that we track are shown in blue in Fig.~\ref{Fig:ion_dynamics_more}d. The right bunch moves ahead of the expanding ion front, so we will refer to it as an \textit{early bunch}. The left bunch moves through the channel filled with ions, so we will refer to it as a \textit{late bunch}. The electrons are grouped in bunches because they can only be periodically injected by the transverse laser electric field during a specific part of each laser cycle. The simulations discussed in Sec.~\ref{Sec-4} were set up such that the ion mobility only impacts the electron acceleration. Electron injection remains unaffected by changes in the ion mass-to-charge ratio in the acceleration region. We can therefore pick out and track the same two bunches in a simulation with immobile ions ($\mu = \infty$). This allows us to clearly identify the feedback that the expanding ions have on the dynamics of accelerated electrons.

Figure~\ref{Fig:early_bunch_inst} shows the time evolution of the early bunch. The bottom row is for the case with $\mu = 2$ in the acceleration region, whereas the top row is for the immobile case ($\mu = \infty$). The electron energies as well as the location and structure of the bunch are essentially the same in both cases. This agrees well with our expectation that the ion mobility should not influence the early bunch. 

The electrons in this bunch are moving predominately in the forward direction. Indeed, the transverse spread over $\Delta y \approx 2R \approx 5.4$ $\mu$m occurs as the bunch travels roughly $\Delta x \approx 60$ $\mu$m in the forward direction. The electrons are highly relativistic, so we can estimate the characteristic transverse velocity as $v_{\perp} \approx c \Delta y / \Delta x \approx 0.09 c$, whereas the longitudinal velocity $v_{\parallel}$ is very close to the speed of light $c$. Since $v_{\perp} \ll v_{\parallel}$, the work done by the transverse laser electric field is strongly suppressed for the early bunch [see Eq.~(\ref{W_x and W_y})]. It is worth pointing out that the transverse spread decreases as electron energy increases which means that the contribution from $E_{\perp}$ is even smaller for higher energy electrons.

The predominantly longitudinal motion of the early bunch is sustained because the transverse forces induced by transverse electric and magnetic fields almost compensate each other. The transverse force acting on an ultra-relativistic forward moving electron, $v_x \rightarrow c$, is
\begin{equation} \label{F_y}
    F_y = -|e| \left( E_y - \frac{v_x}{c} B_z \right) \approx -|e| \mathscr{E} E_0,
\end{equation}
where
\begin{equation} \label{E_norm}
    \mathscr{E} \equiv (E_y - B_z) / E_0.
\end{equation}
We explicitly assume that the $x$-component of the magnetic field is negligible in our simulations, which is indeed the case for the laser field and for the field generated by the plasma. The magnitude of $\mathscr{E}$ then quantifies how well the transverse forces compensate each other, with $\mathscr{E}$ of the order of unity corresponding to a lack of compensation. The spatial profile of $\mathscr{E}$ is shown in Fig.~\ref{Fig:early_bunch_inst} with the purple-green color-scheme. These plots confirm that there is indeed a significant compensation inside the channel, with $\mathscr{E} \ll 0.1$. 

In order for $\mathscr{E}$ to be small inside the channel, the condition $E_y \approx B_z$ must be satisfied for both (1) the oscillating fields and (2) the quasi-static fields. This distinction is important because the oscillating fields are dominated by the fields of the laser pulse whereas the quasi-static fields are generated by the plasma. For a plane wave $E_y = B_z$ exactly, but in our case the field structure is similar to that of a TM-mode in a wave-guide. For a TM-mode, we have $|E_y| < |B_z|$\cite{jackson1975electrodynamics}, so that
\begin{equation}
    |E_y - B_z| / E_0 \propto \lambda^2 / (2R)^2.
\end{equation}
A significant compensation can then still take place for the fields of the laser pulse, provided that the channel width exceeds the laser wavelength.  

Even though the amplitude of the oscillating fields can be much greater than the amplitude of the quasi-static fields, their contribution to $\mathscr{E}$ is greatly diminished because of the discussed compensation and that is why the role of the quasi-static fields becomes important. In order to determine the quasi-static fields, the values of $E_y$ and $B_z$ in our simulations are time-averaged over four laser periods. The resulting quantities are denoted with an over-line, e.g. $\overline{E}_y$ and $\overline{B}_z$. In our previous publication, we presented a detailed analysis of the quasi-static fields generated by the accelerated electrons in the regime where the ion expansion is negligible~\cite{Gong_hollow_channel}. The electrons generate both electric and magnetic fields. The amplitude of the magnetic field becomes comparable to the amplitude of the electric field for forward-moving ultra-relativistic electrons, such that $|\overline{E}_y - \overline{B}_z| \ll |\overline{E}_y|$. This compensation is the very reason why the electrons can move almost directly forward while being effectively accelerated by the longitudinal electric field of the laser pulse~\cite{Gong_hollow_channel}. Figure~\ref{Fig:early_bunch_av} shows that the same compensation takes place for the early bunch, because the ion motion is insignificant at this stage.

We now turn our attention to the time evolution of the late bunch. The bottom row in Fig.~\ref{Fig:late_bunch_inst} shows snapshots of the late bunch in a channel with mobile ions ($\mu = 2$) in the acceleration region. The top row in Fig.~\ref{Fig:late_bunch_inst} shows the same bunch moving through a channel with immobile ions ($\mu = \infty$). The presence of the ions in the channel dramatically changes the dynamics of the bunch (Figs.~\ref{Fig:late_bunch_inst}e and \ref{Fig:late_bunch_inst}f) compared to the case with immobile ions (Figs.~\ref{Fig:late_bunch_inst}b and \ref{Fig:late_bunch_inst}c). The electrons in the bunch gain less energy and they also have a much higher transverse velocity. 

The changes in the dynamics of the bunch are caused by an increase in the transverse force which is apparent in plots $\mathscr{E}$ in Figs.~\ref{Fig:late_bunch_inst}e and \ref{Fig:late_bunch_inst}f. The force causes the electrons to move towards the central axis of the channel, resulting in an increase in their transverse velocity. This change increases the work done by the transverse electric field $E_{\perp}$ while also reducing the contribution by $E_{\parallel}$. 

The reduction in $\mathrm{W}_{\parallel}$ can be understood by considering the change in the longitudinal velocity $v_{\parallel}$. The electrons are highly relativistic, so their total velocity still remains close to the speed of light as they accelerate towards the axis. This means that the increase in $v_{\perp}$ comes at the expense of $v_{\parallel}$. As $v_{\parallel}$ drops, so does the work performed by $E_{\parallel}$. Moreover, the drop in $v_{\parallel}$ increases the relative velocity between the electrons and the wave-fronts. The amount of time the electrons can spend in the favorable phase will decrease which in turn limits the energy they can gain from $E_{\parallel}$.

\begin{figure}[htb]
    \centering
    \includegraphics[width=0.99\columnwidth]{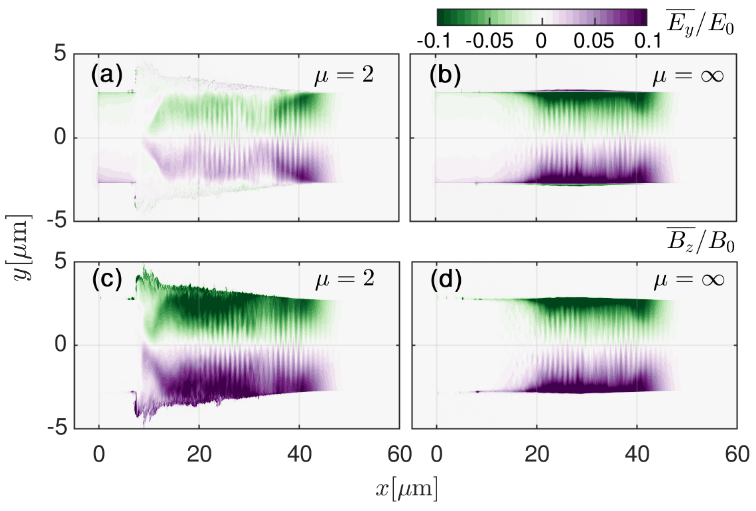}
    \caption{Comparison of time-averaged electric (a and b) and magnetic (c and d) fields for channels with immobile ions, $\mu = \infty$, (right) and mobile ions, $\mu = 2$, (left) in the accelerating region.}
    \label{field_comparison}
\end{figure}

Our last step is to determine the cause of the uncompensated force acting on the late bunch. Figure~\ref{Fig:late_bunch_inst} shows the profile of $\mathscr{E}$, which represents the instantaneous transverse force on a forward-moving electron. The apparent lack of compensation between $E_y$ and $B_z$ may be caused by either the oscillating or quasi-static fields.
Figure~\ref{Fig:late_bunch_inst} shows the profile of $\overline{\mathscr{E}}$, which is the value of $\mathscr{E}$ averaged over four laser periods at a fixed location. The amplitude of $\overline{\mathscr{E}}$ is very similar to the amplitude of $\mathscr{E}$ at $t = 50$ fs and $t = 100$ fs [bottom rows in Figs.~\ref{Fig:late_bunch_av} and \ref{Fig:late_bunch_inst}]. We can therefore conclude that the force causing the transverse motion of the late electron bunch towards the central axis is a manifestation of uncompensated quasi-static electric and magnetic fields generated by electrons and ions.

The profiles of quasi-static electric and magnetic fields at $t = 110$ fs are shown in Fig.~\ref{field_comparison}. As already discussed, the electric and magnetic fields in the immobile case ($\mu = \infty$) are nearly identical (see Figs.~\ref{field_comparison}b and \ref{field_comparison}d). These fields are generated exclusively by the electrons moving through the channel. In the case of $\mu = 2$, the ion expansion provides a compensating positive charge inside the channel and leads to a significant reduction of the transverse quasi-static electric field in Fig.~\ref{field_comparison}a as compared to Fig.~\ref{field_comparison}b.
The expansion takes time to manifest itself, which explains why the electric field remains strong for $x > 40$~$\mu$m in this snapshot. 

The quasi-static magnetic field is generated by the electron current so it is not directly impacted by the ion motion. However, the ions do have an impact on the spatial profile of the magnetic field. The reduction of the electric field leads to an uncompensated transverse force that causes the electrons carrying the current to spread across the channel rather than remain concentrated near the channel walls. As a result, the magnetic field becomes volumetrically distributed when the ion expansion is significant, as seen in Fig.~\ref{field_comparison}c. 

The results presented in this section conclusively demonstrate that the ion expansion reduces the quasi-static transverse electric field in the channel. This leads to an uncompensated transverse force directed towards the central axis of the channel. The  resulting transverse oscillations of the electrons increase the work done by $E_{\perp}$ while simultaneously reducing the work done by $E_{\parallel}$ because of the decrease in longitudinal electron velocity.

\textcolor{black}{\section{Assessment of the impact from a laser pre-pulse}} \label{Sec-prepulse}

\begin{figure*}[htb]
\centering
    \includegraphics[width=1.0\textwidth,trim={0.0cm 0.1cm 0 0.1cm},clip]{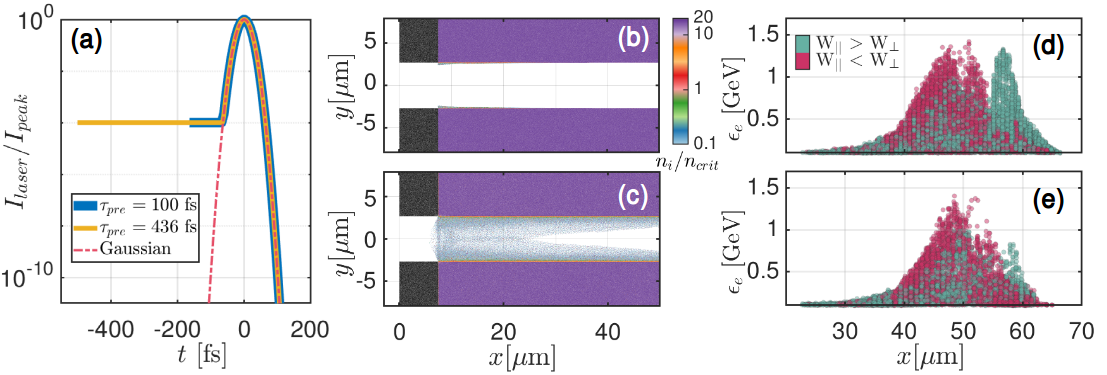}
\caption{\textcolor{black}{Impact of the pre-pulse duration on ion pre-expansion and electron acceleration in the main pulse. (a) Temporal profiles of laser intensity for two different pre-pulse durations. (b) and (c) Snapshots of the ion density at the start of the intensity ramp-up associated with the main pulse ($t \approx -64$~fs). (d) and (e) Snapshot of accelerated electrons at $t = 171$~fs. Panels (b) and (d) correspond to the shorter pre-pulse, whereas panels (c) and (e) correspond to the longer pre-pulse.}}
\label{fig:prepulse}
\end{figure*}

\textcolor{black}{We have so far considered a laser pulse with a Gaussian time profile, which is a deliberate simplification. Experimentally used laser pulses often contain a pre-pulse. In the context of the considered setup, a pre-pulse can have a profound impact on electron acceleration by inducing ion expansion prior to the arrival of the main laser pulse. In what follows, we show how to assess the role of a given pre-pulse using the already obtained results.}

\textcolor{black}{In order to determine how a pre-pulse impacts the ion density profile in the channel, we treat the pre-pulse as a separate laser pulse that temporally overlaps with the main pulse. Two examples of laser pulses containing a pre-pulse are shown in Fig.~\ref{fig:prepulse}a. At $t > -64$~fs, the laser intensity has a Gaussian profile, so, for convenience, we refer to $t \approx -64$~fs as the ``beginning'' of the main pulse. It is evident from Fig.~\ref{fig:prepulse}a that the prepulse in both cases can indeed be viewed as a standalone laser pulse with its own intensity and duration that precedes the main pulse.} 

\textcolor{black}{We next apply the approach developed in Sec.~\ref{Sec-2} to re-examine the already considered laser-channel configuration with an additional pre-pulse. We consider two sub-ps prepulses shown in Fig.~\ref{fig:prepulse}a that differ in duration. The pre-pulse intensity is four orders of magnitude lower than the peak intensity of the main pulse, so we have $I = 5.0\times10^{18}$ W/cm$^2$ or, equivalently, $a_0 \approx 2$ in both pre-pulses. It follows from Eq.~(\ref{Delta t - 2}) that the estimated ion travel time at this pre-pulse amplitude is approximately 340~fs. Our model thus predicts that the impact of the shorter pre-pulse should be negligible, whereas we should expect for the longer pre-pulse to cause significant ion expansion prior to the beginning of the main pulse.}

\textcolor{black}{To check the predictions of our model, we carried out PIC simulations for the two laser pulses shown in Fig.~\ref{fig:prepulse}a. The snapshots of the ion density at the beginning of the main pulse ($t \approx - 64$~fs) are shown in Figs.~\ref{fig:prepulse}b and ~\ref{fig:prepulse}c. They quantify the impact of the pre-pulse in both cases. In agreement with our estimates, the ions fill the channel only in the case of the longer pre-pulse. The impact of the ion pre-expansion on electron acceleration in the main pulse is illustrated in Figs.~\ref{fig:prepulse}d and~\ref{fig:prepulse}e. In the case of a shorter pre-pulse, there are two distinct groups of electrons. The leading group is accelerated primarily by the longitudinal electric field, as in the simulation without the pre-pulse. In the case of a longer pre-pulse, the longitudinally accelerated group almost disappears. This is a signature of a channel filled with ions, which indicates that there are enough ions in the channel to impact the acceleration of even the very first bunch produced by the main laser pulse.}


\textcolor{black}{These results show that our approach can be extended to evaluate the impact of a known pre-pulse by treating it as a separate laser pulse, which demonstrates the predictive capability of the criterion provided in Eq. (\ref{Delta t - 2}). Typically, pre-pulse simulations are time-consuming, so the derived criterion can be used to determine whether one needs to include the pre-pulse into the PIC simulation or not.}


\section{Influence of ion mobility on x-ray and gamma-ray emission} \label{Sec-6}

\begin{figure*}[htb]
	\begin{subfigure}[c]{0.5\linewidth}{
    \includegraphics[width=0.98\textwidth,trim={0.0cm 1.cm 0 0.5cm},clip]{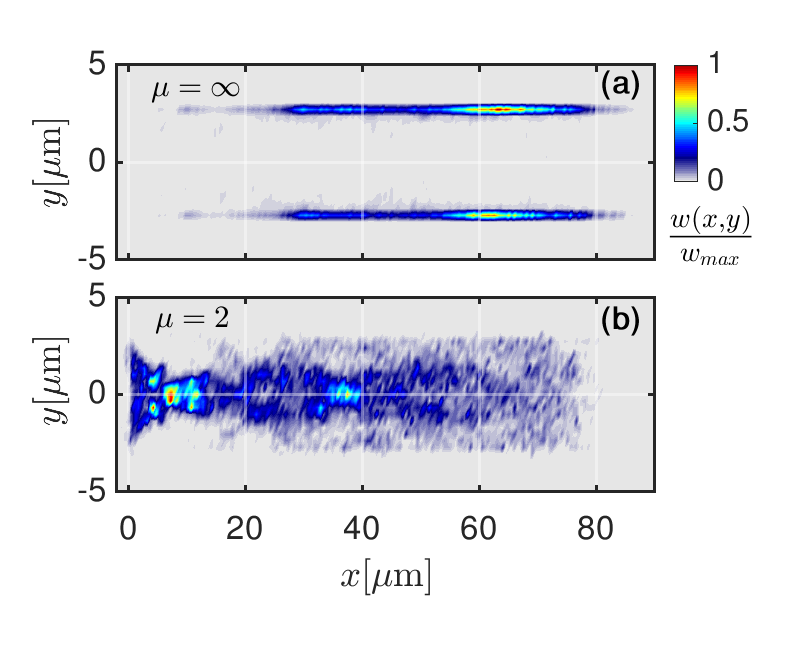}}
    \end{subfigure}
	\begin{subfigure}[c]{0.45\linewidth}{
    \includegraphics[width=0.98\textwidth]{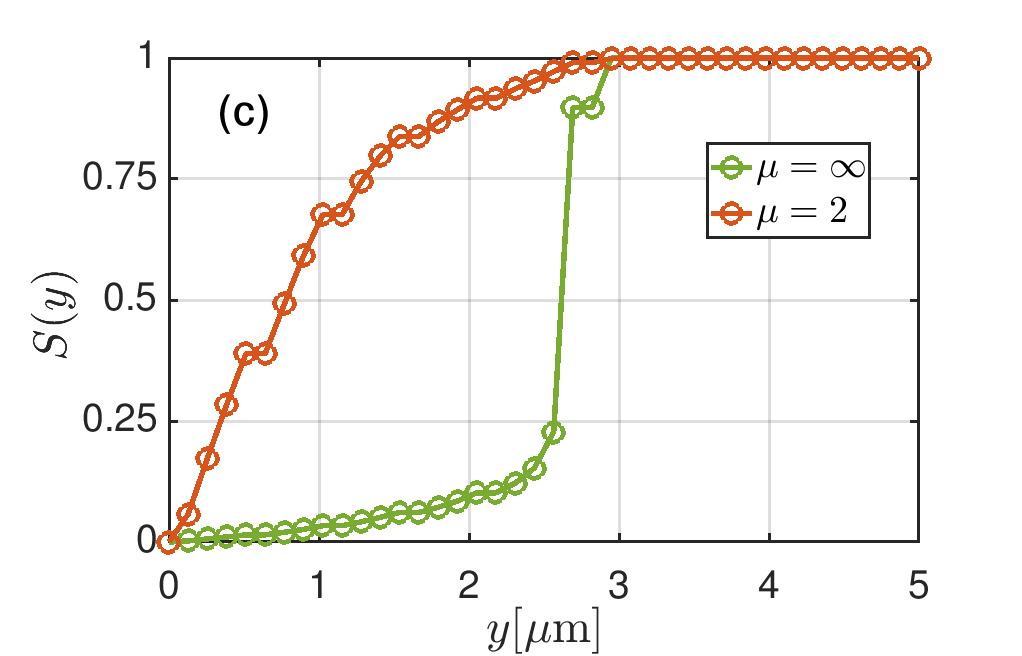}}
    \end{subfigure}
    \caption{Snapshots of the time-integrated gamma-ray emission at $t = 270$ fs for photons with energies above 10 MeV. Panels (a) and (b) are the profiles of the emitted energy in channels with immobile ($\mu = \infty$) and mobile $(\mu = 2)$ ions. Panel (c) is the fraction of the total energy emitted between the central axis and the transverse location specified by $y$ [see Eq.~(\ref{S})].}    \label{Emission}
\end{figure*}

Our attention in Sections \ref{Sec-3} - \ref{Sec-5} was focused on the impact of the ion dynamics on laser-driven electron acceleration and generation of energetic electrons in the GeV energy range. These high-energy electrons have the potential to emit energetic x-ray photons and even gamma-rays when undergoing acceleration. In this section, we analyze how the photon emission changes as a result of the ion expansion.

An electron accelerated by electric and magnetic fields emits electromagnetic radiation. The emitted power, $P$, \textcolor{black}{of the corresponding synchrotron emission} is determined by the electron acceleration in an instantaneous rest frame. This acceleration is often quantified using a dimensionless parameter~\cite{landau1971classical,duclous2010monte,ridgers2014modelling}
\begin{equation}
    \eta \equiv \frac{\gamma}{E_s} \sqrt{\bigg({\bf{E}} + \frac{1}{c} [{\bf{v \times B}}] \bigg)^2 - \frac{1}{c^2}({\bf{E \cdot v}})^2 },
    \label{for_eta}
\end{equation}
where $\gamma$ and $v$ are the relativistic factor and velocity of the electron and $E_S \approx 1.3 \times 10^{18}$ V/m is the Schwinger limit. The radiated power scales as $P \propto \eta^2$. If $\eta \ll 1$, then the emission can be treated as a continuous process~\cite{di2012extremely}. In the regimes considered in this work, this condition is not satisfied and the emission process has to be treated as emission of individual photons~\cite{ridgers2014modelling}. We simulate the photon emission using the module implemented in the particle-in-cell code EPOCH~\cite{Epoch}.


It is instructive to examine the parameter $\eta$ for an ultra-relativistic forward moving electron with $v_x \approx c$ and $v_y = 0$. The expression given by Eq.~(\ref{for_eta}) reduces to
\begin{eqnarray}
    \eta &\approx& \frac{\gamma E_0}{E_s} \left[ \left( \mathscr{E} + \frac{1}{2\gamma^2} \frac{B_z}{E_0} \right)^2 + \frac{1}{\gamma^2} \frac{E_x^2}{E_0^2} \right]^{1/2},
    \label{eta_forward}
\end{eqnarray}
where we explicitly took into account that the magnetic field only has the $B_z$ component, while the electric field has $E_x$ and $E_y$ components. The second and third terms are strongly suppressed because of the $1/\gamma^2$ multipliers. If we neglect them, we find that
\begin{equation}
    \eta \approx \left. \gamma |\mathscr{E}| E_0 \right/ E_s.
    \label{eta_forward2}
\end{equation}
Therefore, $\eta$ and the resulting emitted power are determined in this case exclusively by the combination of the fields that we defined as $\mathscr{E}$ in Eq.~(\ref{E_norm}).

Even though the derived expression is approximate, it enables us to draw preliminary conclusions based on the numerical results for $\mathscr{E}$ obtained earlier. As shown in Fig.~\ref{Fig:late_bunch_inst}, the ion expansion significantly enhances the amplitude of $\mathscr{E}$ inside the channel when compared to the immobile case ($\mu = \infty$). This should enhance the emission by the electron bunches inside the channel for $\mu = 2$. In the immobile case ($\mu = \infty$), $|\mathscr{E}|$ peaks at the wall as a result of an uncompensated electric field that deflects energetic electrons back into the channel once they reach the wall~\cite{Gong_hollow_channel}. We should therefore expect that the emission in a channel with immobile ions should be primarily concentrated at the walls.

These qualitative conclusions are in good agreement with the calculated emission patterns shown in Fig.~\ref{Emission}. Figure~\ref{Emission}a corresponds to an immobile case, whereas Figure~\ref{Emission}b corresponds to a mobile case with $\mu = 2$. The color represents the time-integrated energy emitted in each of the grid cells. Specifically, the plotted quantity is obtained by time integrating the energy that is emitted in a given cell centered at $(x,y)$ for $t < 270$ fs:
\begin{equation} \label{w-def}
    w (x,y) = \sum_{\epsilon_{\gamma} > \epsilon_c} \epsilon_{\gamma} N_{\gamma} (x,y),
\end{equation}
where $\epsilon_{\gamma}$ is the photon energy, $N_{\gamma} (x,y)$ is the number of photons emitted with this energy, and $\epsilon_{c} = 10$ MeV is the cutoff photon energy. The electrons continue to emit at $t > 270$ fs, but our conclusions remain unaffected. 
 


Figure~\ref{Emission}c shows how the energy emission is distributed across the channel. We calculate this quantity by time integrating the energy that is emitted at a given transverse position $y$ for $t < 270$ fs:
\begin{equation}
    \epsilon (y) = \int_{-\infty}^{+ \infty} w(x,y) dx,
\end{equation}
where $w(x,y)$ is defined by Eq.~(\ref{w-def}). The quantity $S(y)$ that is plotted in Fig.~\ref{Emission}c shows what fraction of the total emitted energy is emitted between the central axis and the transverse location specified by $y$:
\begin{equation} \label{S}
   S(y) \equiv \left.  \int_{0}^{y} \epsilon (y') dy' \right/ \int_{0}^{\infty} \epsilon (y') dy'.
\end{equation}
In the mobile case, about $85\%$ of the energy is emitted inside the channel at $|y| < 2$ $\mu$m. In contrast to that, only 10\% of the energy is emitted at $|y| < 2$ $\mu$m if the ions are immobile.


The ion expansion not only changes the emission from being concentrated at the walls to being volumetrically distributed, but it also reduces the total emitted energy. We find that for the plotted snapshots the energy emitted by the photons with $\epsilon_{\gamma} > 10$ MeV is almost ten times lower in the mobile case. \textcolor{black}{In the case of $\epsilon_{\gamma} > 100$~keV, the energy emitted in the mobile case is approximately three times lower than that in the immobile case.} There are two factors contributing to these results. As shown in Fig.~\ref{Fig:two_groups}, the electron energies in the mobile case tend to decrease in the considered regime. Then, even though $|\mathscr{E}|$ is a lot higher in the mobile case inside the channel, it is still lower than the peak of $|\mathscr{E}|$ at the wall in the immobile case. 

Our results then indicate that the ion expansion has a profound impact on the gamma-ray emission in laser-irradiated hollow channels. The emission becomes volumetrically distributed, while the total emitted energy is reduced.



\section{Summary and Discussion} \label{Sec-7}

In this work, we have examined the impact that the ion dynamics  has on laser-driven  electron  acceleration  in  an initially empty channel. The negative charge of the accelerated electrons inside the channel generates a quasi-static transverse electric field that causes gradual ion expansion into the channel. There are two distinct populations of accelerated electrons:  those that accelerate  ahead  of  the  expanding  ion  front  and  those that accelerate in the presence of the ions in the channel. The ions significantly reduce the quasi-static transverse electric field, so that there is an uncompensated transverse force from the quasi-static magnetic field that pulls electrons towards the central axis of the channel. The resulting transverse oscillations of the electrons increase the work done by $E_{\perp}$ while simultaneously reducing the work done by $E_{\parallel}$. We also found that the ion expansion has a profound impact on the gamma-ray emission in laser-irradiated hollow channels. The emission becomes volumetrically distributed, while the total emitted energy is reduced.

One of the main results of this work is the criterion provided by Eq.~(\ref{Delta t - 2}) that specifies the time interval $\Delta t$ needed for the ion expansion to manifest itself after the start of the interaction with the laser pulse at a given location. It should be viewed as an indicator for the laser pulse duration $\Delta \tau$. Since $\Delta t$ scales as $1/\sqrt{a_0}$, it imposes a stringent requirement on the duration of laser pulses with $a_0 > 100$. If $\Delta \tau > \Delta t$, then most of the accelerated electrons perform strong transverse oscillations and only the front part of the laser pulse is able to produce highly collimated electrons.

In this work, we treat the normalized mass-to-charge ratio $\mu$ as an input parameter in order to systematically investigate its impact. However, $\mu$ must be determined by including ionization processes in the simulations to provide quantitative predictions. The normalized mass-to-charge ratio $\mu$ decreases with the increase of $a_0$ due to field ionization and the change in $\mu$ can be significant for $a_0 > 100$. For example, we have performed a simulation with a gold channel where the ionization process was included. The field ionization at $a_0 \approx 190$ has resulted in a relatively low mass-to-charge ratio of $\mu \approx 2.9$. The main implication of this result is that simulations with a prescribed value of $\mu \gg 2$ would severely underestimate the impact of the ions at ultra-high laser amplitudes ($a_0 > 100$) even for laser pulses that are only tens of femtoseconds long.

The ion dynamics with an appropriate value of $\mu$ is particularly important when calculating the gamma-ray yield and the efficiency of the energy conversion into multi-MeV ions. The emission takes place in a very thin layer near the wall of the channel only if the ions are very heavy. The ion expansion changes this patterns and reduces the energy conversion efficiency. Therefore, simulations with $\mu \approx 2$ are likely to provide a realistic assessment of the photon emission at $a_0 > 100$ if the inclusion of ionization processes is not feasible. 

\textcolor{black}{The criterion provided by Eq.~(\ref{Delta t - 2}) can be applied to evaluate the impact of a given pre-pulse on ion expansion without performing additional PIC simulations. The ion expansion during the pre-pulse is important because it can alter the field configuration in the channel such that even the dynamics of the very first electron bunch produced by the main pulse is affected. The resulting transverse oscillations of electrons across the channel tend to smear out the localized bunches, which can have far-reaching consequences. For example, the existence of these bunches is necessary for generation of attosecond flashes~\cite{lecz2019attosecond}. We can thus conclude that the criterion~(\ref{Delta t - 2}) can also provide a useful guideline for this application.}

Finally, it is important to stress that the ion dynamics changes the electron acceleration mechanism at $a_0 \gg 1$ from being dominated by $E_{\parallel}$ to being dominated by $E_{\perp}$. In our simulations, this led to a decrease in the maximum ion energy. However, the transverse electron oscillations enable a new energy enhancement mechanism that is linked to electron deflections by the quasi-static magnetic field~\cite{gong2018_FSSA}. Optimization of this mechanism in the context of initially empty channels should allow one to mitigate the negative impact of the ion expansion on the electron energy gain.

\section*{Acknowledgements}

This research was supported by the National Science Foundation under Grant No. 1632777 and AFOSR under grant No. FA9550-17-1-0382. Particle-in-cell simulations were performed using EPOCH~\cite{Epoch}, developed under UK EPSRC grants EP/G054940, EP/G055165, and EP/G056803. High performance computing resources were provided by Texas
Advanced Computing Center (TACC) at The University of Texas at Austin and by the Extreme Science and Engineering Discovery Environment (XSEDE), which is supported by National Science Foundation grant number ACI-1548562. We appreciate the data collaboration supported by the SeedMe2 project (http://dibbs.seedme.org)\cite{amit:2017:SDS:3093338.3104153}.


\section*{References}

\bibliographystyle{apsrev4-1}
\input{output.bbl}

\end{document}

%% file: output.bbl
%